\documentclass[12pt]{article}

\usepackage{amssymb}
\usepackage{amsmath}
\usepackage{latexsym}
\usepackage{epsfig}
\usepackage{graphicx}
\usepackage{subfigure}
\usepackage{wasysym}
\usepackage{threeparttable}
\usepackage{booktabs}
\usepackage[round]{natbib}
\usepackage{color}
\usepackage{epstopdf}
\usepackage{bm}
\usepackage{tabularx}
\usepackage{verbatim}
\usepackage{hyperref}
\usepackage{float}
\usepackage[ruled,vlined]{algorithm2e}
\usepackage{ulem}
\usepackage{listings}
\usepackage{color}
\usepackage[usenames,dvipsnames]{xcolor}

\def\log{\hbox{log}}

\def\boxit#1{\vbox{\hrule\hbox{\vrule\kern6pt
          \vbox{\kern6pt#1\kern6pt}\kern6pt\vrule}\hrule}}

\def\bse{\begin{eqnarray*}}
\def\ese{\end{eqnarray*}}
\def\be{\begin{eqnarray}}
\def\ee{\end{eqnarray}}
\def\bq{\begin{equation}}
\def\eq{\end{equation}}
\def\bse{\begin{eqnarray*}}
\def\ese{\end{eqnarray*}}

\begin{document}
\thispagestyle{empty}
\baselineskip=28pt

\begin{center}
{\LARGE{\bf A Look into the Problem of Preferential Sampling from the Lens of Survey Statistics}}

\end{center}

\baselineskip=12pt

\vskip 2mm
\begin{center}
Daniel Vedensky\footnote{(\baselineskip=10pt to whom correspondence should be addressed)
Department of Statistics, University of Missouri,
146 Middlebush Hall, Columbia, MO 65211-6100, dvedensky@mail.missouri.edu},
Paul A. Parker\footnote{\baselineskip=10pt
Department of Statistics, University of California Santa Cruz,
1156 High St, Santa Cruz, CA 95064, paulparker@ucsc.edu},
   and Scott H. Holan\footnote{\baselineskip=10pt Department of Statistics, University of Missouri,
146 Middlebush Hall, Columbia, MO 65211-6100, holans@missouri.edu}\,\footnote{\baselineskip=10pt U.S. Census Bureau, 4600 Silver Hill Road, Washington, D.C. 20233-9100, scott.holan@census.gov},
\\
\end{center}
\vskip 4mm

\baselineskip=12pt

\begin{center}
{\bf Abstract}
\end{center}

An evolving problem in the field of spatial and ecological statistics is that of preferential sampling, where biases may be present due to a relationship between sample data locations and a response of interest. This field of research bears a striking resemblance to the longstanding problem of informative sampling within survey methodology, although with some important distinctions. With the goal of promoting collaborative effort within and between these two problem domains, we make comparisons and contrasts between the two problem statements. Specifically, we review many of the solutions available to address each of these problems, noting the important differences in modeling techniques. Additionally, we construct a series of simulation studies to examine some of the methods available for preferential sampling, as well as a comparison analyzing heavy metal biomonitoring data.

\baselineskip=12pt
\par\vfill\noindent
{\bf Keywords:}  Bias correction, Official Statistics, Pseudolikelihood, Spatial.
\par\medskip\noindent

\clearpage\pagebreak\newpage \pagenumbering{arabic}
\baselineskip=24pt

\section{Introduction}

Rooted in survey methodology, issues surrounding {\it informative sampling} (IS) have experienced extensive research in recent years. The problem arises when the probability of selecting a unit to the sample is correlated with the response \citep{pfe07}. If left unaddressed when specifying  a statistical model, this issue may introduce substantial biases. Although many solutions have been proposed for this problem, it remains an extremely active area of ongoing research. One reason for the continued interest is that official statistical agencies disseminate tabulations from key surveys, such as the American Community Survey, that support the distribution of billions of dollars in funds annually (\url{https://www.census.gov/programs-surveys/acs/about.html}). Consequently, adequate modeling of survey data, that properly accounts for the survey design, is increasingly important.

In contrast, more recently, the problem of \textit{preferential sampling} (PS) has been studied in the context of ecological and spatial statistics. Specifically, PS arises when there is dependence between the process or response that is being modeled and the process that gives rise to the data locations \citep{Diggle2010}. Again, it has been observed that substantial bias may be introduced when PS is unaccounted for in the model. There are many important applications that use PS data, ranging from species distribution modeling \citep{pennino2019accounting}, to pollution monitoring \citep{Zidek2014}, to sports analytics \citep{jiao2019bayesian}. \cite{Diggle2010} introduced the foundational model for PS through a shared process for the data locations and responses. Outside of model development, \cite{dinsdale2019methods} propose alternative Monte Carlo estimates to those used by \cite{Diggle2010}, whereas \cite{da2015optimal} consider the problem of optimal sampling design under effects of PS. Finally, \cite{watson2021perceptron} develops a test to detect PS.

These two problems have very similar definitions, yet there are some key differences. First, the sample domain is different within each problem. PS occurs in geostatistical settings, where locations are sampled from a continuous, often multi-dimensional, domain. Conversely, surveys sample individual units (e.g., people, households, establishements, etc.) from a finite population. Another important distinction is the scope of the data included in a sample. Survey data is most often accompanied by a survey weight that is inversely proportional to the unit probability of selection. These selection probabilities are known for all sampled units based on the survey design and sampling frame. However, geostatistical data are rarely accompanied by probabilities of selection. Much of the literature around PS was developed for ecological applications. However, in some cases, data for ecological applications may come from a known sample design (e.g. see \cite{irvine2018occupancy}).

There is a vast literature surrounding various methodologies that may be used to account for IS with unit-level survey data. \cite{parker2019unit} give an overview of these potential approaches. One popular solution is to use an exponentially weighted pseudo-likelihood \citep{bin83, ski89}. Other approaches include nonlinear regression on the survey weights \citep{si15} and inferring a population level model through specification of a model for the weights \citep{pfe07}. Each of these methods relies on the reported survey weights to account for the survey design.

Despite the similarities between these two problems, the development of methodology for PS has happened mostly independently from the IS literature. This is likely due to the fact that IS solutions rely on reported survey weights, which are not typically available in PS applications.
The most common approach to handling PS is to model the sampled locations as a point process and then use a shared latent process for the point process model and the response model \citep{Diggle2010, Pati2011}. However, \cite{Zidek2014} take an approach that is similar to the pseudo-likelihood method that is often used in the IS literature. Their solution is to exponentially weight the likelihood by the inverse of the estimated probabilities of selection. This indicates that although the comparison has not been explored explicitly, it is likely that researchers in the PS field have, to some extent, recognized the similarity to IS.

Our goal in this article is to make an explicit connection between PS and IS. We argue that despite some differences in the problem statements, these are both manifestations of a general problem for which similar methodologies may be used. It is our hope that this will help to foster future research both within and across these areas. 

The remainder of this paper proceeds as follows. In Section~\ref{sec: prob} we introduce notation and give formal problem statements for both IS and PS and classify solutions to these problems into three general groups. Then, in Section~\ref{sec: cov}, we explore methods for both PS and IS that rely on the use of covariates to adjust for the sampling design. In Section~\ref{sec: reg} we investigate methods that use nonlinear regression on the selection probabilities, while in Section~\ref{sec: PL} we explore exponentially weighted likelihood approaches. We conduct a simulation study to compare a subset of methods for PS in Section~\ref{sec: sim}. Note that we do not conduct simulations for IS, as this was previously done by \cite{parker2019unit}. We also illustrate a subset of methods for PS in Section~\ref{sec: examp} using the heavy metal biomonitoring data discussed in \cite{Diggle2010}. Finally, we provide concluding discussion in Section~\ref{sec: disc}.

\section{Problem Statements}\label{sec: prob}

In order to describe the similarity between IS and PS, while still recognizing the key differences, we use a similar although slightly different notation for each problem. In general, we represent the response data, or observations, with $Z.$ Under IS problems, these data are sampled from a finite population, $\mathcal{U}={1,\ldots,N}.$ The sample, of size $n$, is denoted $\mathcal{S} \subset \mathcal{U}.$ Typically, survey units are observed in one of $d=1,\ldots,D$ geographic areas or domains (e.g., census tract, county, etc.). Thus, we index the data for a given observational unit in domain $d$ as $Z_{jd}$, where $j \in \mathcal{S}.$ In contrast to this, preferentially sampled data are typically sampled from a continuous domain, $\mathcal{D}$ (often $\mathcal{D} \subset \mathcal{R}^2$). We denote these spatially referenced data as $Z(s),$ where $s \in \mathcal{D}.$

We also define a latent process as either $Y_{jd}$ in the context of survey data or $Y(s)$ in the context of data sampled from a continuous spatial domain. The latent process may be a function of a vector of covariates ($\bm{x}_{jd}$ or $\bm{x}(s)$) as well as spatially correlated random effects ($\bm{\eta}_d$ or $\bm{\eta}(s)$). Note that in the context of survey data, units are nested within geographic areas and, thus, can be mapped to their corresponding area-level random effects through the use of incidence vectors.

One unique aspect of survey datasets is that they are often accompanied by unit survey weights, $w_{jd}.$ These weights are typically assumed to be the inverse probabilities of selection, $w_{jd}=1/p_{jd}$; however, they may include adjustments for various reasons such as nonresponse. The selection probabilities, and thus the weights, are usually known from the survey design. When data are sampled from a continuous domain, as is the case with most PS problems, there is no probability of selection. However, for a Poisson point process, the intensity is defined as the limit of the probability of observing a point in a decreasing area, and thus serves as a natural analogy. In some cases, we will construct weights for these continuously sampled data, $w(s)=1/p(s),$ where $p(s)$ is the underlying intensity function evaluated at $s \in \mathcal{D}.$ We note that in this case the intensity must usually be estimated, as it is not defined by a known sampling design.

For geostatistical data, we work with the likelihood
$$
f(\bm{Z}, s) = \int f(\bm{Z}, s, \bm{Y}) d\bm{Y}. 
$$ In most cases, we assume that the distribution of the locations is independent of the process and responses. This is the case under uniform random sampling, and results in the likelihood
\begin{align*}
    \int f(\bm{Z}, s, \bm{Y}) d \bm{Y} &= \int f(\bm{Z}|\bm{Y},s)f(s)f(\bm{Y})d\bm{Y} \\
    & \propto \int f(\bm{Z}|\bm{Y},s) f(\bm{Y}) d\bm{Y}.
\end{align*} Thus, when the data locations are independent from the process and response values, the location model can be ignored. However, when independence is not met (i.e., in the case of PS) either the factorization $\int f(\bm{Z}|s,\bm{Y})f(s|\bm{Y})f(\bm{Y})d\bm{Y}$ or $\int f(\bm{Z}|s,\bm{Y})f(\bm{Y}|s)f(s)d\bm{Y}$ must be used.

Similarly, for survey samples, when the distribution of sample inclusion indicators is independent of the response values and any latent process, the survey design may be ignored. In the case of surveys that exhibit IS, the survey design must be considered in the model.

\section{Use of Informative Covariates in the Model}\label{sec: cov}
One of the most basic ways to correct for an informative design in the survey setting is to include all of the design variables in the model, typically  as covariates. In this case, the response, conditioned on the covariates, is independent of the selection probabilites and inference can be based on this conditional distribution. For instance, including a fixed effect for strata could correct for a stratified design. \citet{little2012} outlines such an approach in a Bayesian setting, in which case it is possible to calculate a posterior predictive distribution for the unsampled population. This predictive distribution can then be used for inference on population quantities of interest.

Another corrective measure from the survey literature that relies on the use of design variables is post-stratification \citep{little1993}. This approach assumes the population contains $C$ categories, or poststratification cells, each with a known population size $N_c$, $c =1,\ldots, C$. Observations within each cell are assumed to be independent and identically distributed. These categories are generally determined by cross-classifications of levels of categorical covariates or of continuous covariates that have been discretized. For example, a post-stratified estimator for the population mean, $\bar{z}_p$, could be calculated as
\[\bar{z}_P = \sum_{c=1}^C \frac{N_c}{N}\bar{z}_{cS},\]
where $\bar{z}_{cS}$ denotes the observed mean response for sampled units in cell $c$.
 
\citet{gelman_little1997} and \citet{park06} combine poststratification with a Bayesian multi-level model, which allows for parameter estimates of cells with no sampled units. For example, with binary data a model of the following form could be used
\begin{align*}
z_{jd} |p_{jd} &\sim Bernoulli(p_{jd})\\
\text{logit}(p_{jd}) &= \bm x'_{jd}\bm\beta\\
\bm\beta &= (\bm\gamma_1,\ldots,\bm\gamma_\ell)\\
\bm\gamma_\ell &\overset{ind}{\sim} N_{m_\ell}(0,\sigma^2_\ell\bm I_{m_\ell}), \ell =1,\ldots,L,
\end{align*}
where $\bm x_{jd}$ is a vector of dummy variables for $L$ categorical predictor variables with $m_\ell$ classes in variable $\ell$. Bayesian inference is performed on this model to get a probability $p_c$ for each cell, $c=1,\ldots, C$. The number of positive responses within cell $c$ is estimated as $N_c p_c,$ and any higher level aggregate estimates can be made by aggregating the corresponding cells. 

A major impediment to using these design-variable based approaches is the fact that all design variables are rarely known. For example, for data users outside of a statistical agency, full design knowledge may not be attainable, and even within a statistical agency, one may not have access to all design variables. In some cases, where all desgin variables are available, their inclusion may complicate the likelihood, or even make it intractable.

Similar difficulties persist in the PS setting, so the approach is not common. However, it is of conceptual interest as a potentially simple and straightforward correction. In the PS case, there may not be a formal sampling design and therefore no design variables in the IS sense. However, \citet{gelfand12} introduce the analogous notion of an ``informative covariate"--- a covariate that is correlated with both the response values and the choice of sampling locations. They take the example of including population density as a covariate in a model that predicts pollution levels. They carry out a simulation study that, in part, assesses how effectively the use of such an informative covariate may reduce bias due to PS. In fact, they find it does little to remedy the bias, let alone make up for the difference between a completely randomized sampling scheme and a PS scheme. 

\citet{conn17} also examine the effect of including informative covariates in models for preferentially sampled data, specifically in the context of animal population models. They, too, conclude that this approach is often inadequate since predictive covariates explain only a small portion of variation present in the data in many contexts. Often, the factor driving location selection may not even be known. However, they do point out that for some sampling designs, there is theoretical justification for collecting more samples in areas where the response is expected to be higher. In this case, sampling according to a covariate and then including it in the model will lead to an ignorable, non-preferential design. They mention post-stratification as a potential way to correct for bias, as well, with the caveat that it may not be clear how to do so when effort is allocated in a subjective manner. Unlike the IS setting, it is also unrealistic to expect knowledge of the population poststratification cell sizes, $N_c$, in the PS case.

\section{Regressing on the Selection Process}\label{sec: reg}
Another commmon and conceptually straightforward approach to alleviating bias in IS or PS situations is to directly specify the dependence relationship between the response and sample selection process in the likelihood. In the field of survey methodology, for example, there is a long history of use of the data model
\begin{equation}\label{eq: infreg}
    Z_{jd} = \bm{x}_{jd}'\bm{\beta} + g(w_{jd}) + \epsilon_{jd},
\end{equation} which attempts to directly adjust for IS through estimation of the function $g(\cdot).$ In the most simple case, \cite{firth98} consider the linear function $g(w_{jd})=a\cdot w_{jd}.$ In practice, the assumption of linearity can be quite restricting, and thus, \cite{zhe03} explore the use of nonlinear models for $g(\cdot)$ via penalized splines. We note that survey data is frequently used for the purpose of small area estimation, in which the area population totals $\sum_{j \in d}g(w_{jd})$ are required to construct estimates. In the case of linear $g(\cdot),$ this is not necessarily restrictive; however, nonlinear specifications of $g(\cdot)$ should be carefully considered.

\cite{si15} use a flexible Gaussian process prior for $g(\cdot).$ They take a unique approach to estimation of population totals by defining poststratification cells according to the unique survey weight values. Through a multinomial data model for the observed cell sizes, they are able to generate predictions of the survey weights for all individuals outside of the sample. \cite{van16} extend this approach to the small area estimation setting by defining poststratification cells as unique combinations of weight and geographic area.

Unlike survey data settings, geostatistical data is not usually accompanied by known sample weights or probabilities. This can introduce further challenges when attempting to model the sampling dependence directly.

\citet{Diggle2010} introduce a foundational approach to handling preferentially sampled data. They assume that the data locations follow a log-Gaussian Cox process \citep{moller1998log}. That is, conditional on an underlying, unobserved, Gaussian process, $Y(s),$ the data locations follow a Poisson point process with intensity
\begin{equation*}\label{eq:diggle_intensity}
p(s) = \hbox{exp}\left\{\alpha +  Y(s) \beta \right\}.
\end{equation*} The response values are simultaneously modeled as
$$
Z(s) = \mu + Y(s) + \epsilon(s),
$$ where $\epsilon(s)$ is independent and identically distributed. Importantly, the latent process $Y(s)$ is shared between the location model and the response model to account for preferentially sampled data.
 Through simulation and analysis of lead pollution data in Galicia, Spain, they show that failure to account for PS leads to substantial prediction bias and underestimated standard errors. Additionally, they show that their approach is able to reduce this bias.

\citet{Pati2011} take a very similar approach within a Bayesian framework. Again, they model the data locations as a log-Gaussian Cox process with intensity
$$
p(s) = \hbox{exp}\left\{Y(s)\right\},
$$ conditional on the latent Gaussian process, $Y(s).$ In doing so, the response model is
\begin{equation}\label{eq: pati}
    Z(s) = \eta(s) + Y(s) \beta + \epsilon(s),
\end{equation} where $\eta(s)$ is an additional spatially correlated Gaussian process that adds flexibility, and again, $\epsilon(s)$ is independent and identically distributed noise. Note that $\eta(s)$ can be defined to include covariate information in the mean structure. In this framework, $\beta$ controls the level of PS in the data, where $\beta=0$ indicates no presence of PS. After conditioning on $Y(s)$ and noting that the intensity, $p(s)$ is analogous to the sample selection probability in the discrete case, (\ref{eq: pati}) becomes reminiscent of the IS Model (\ref{eq: infreg}), with $g\left(w(s)\right) = \hbox{log}\left(\frac{1}{w(s)}\right)\beta$. Both models take the approach of regressing on a function of the selection process in order to account for selection bias. \cite{grantham2018spatial} embed this approach into a deeper hierarchical model in order to account for informative missingness of geostatistical data.

Both \cite{Diggle2010} and \cite{Pati2011} only consider the case of a continuous (Gaussian) response variable over a continuous spatial domain. In an effort to expand the applicability of these methods, \cite{conn17} extend to the case of count data on a discrete (areal) domain, as well as the case where only some parts of the domain are sampled preferentially. For the discrete domain, data locations are observed over a finite grid space, eliminating the need for the point process model. This results in a true probability of selection, although one that is still typically unobserved in practice. Similarly, \cite{gelfand19} consider the case of presence/absence as well as presence only data through a shared process method and \cite{pennino2019accounting} consider abundance data under PS schemes.

The general approach of regressing on the selection process through a shared latent process has dominated much of the literature in PS. However, this general approach makes up comparatively less of the literature in the IS world. This may be in part due to the fact that survey data under informative sample designs are typically accompanied by a survey weight. Considering the weights as fixed and known could allow for a broader class of methods than the case where weights or selection probabilities must be modeled, as is the case with the shared process models discussed herein.

One limitation of (\ref{eq: pati}) is the assumption of linearity between $Z(s)$ and $Y(s).$ In scenarios where data is frequently observed in locations with both high and low expected response values, this linear assumption is flawed. Yet, estimation of a nonlinear function $g\left(Y(s)\right)$ can be challenging when $Y(s)$ is a stochastic process itself. In contrast, the survey literature frequently considers nonlinear $g(w_{jd}),$ allowable in part due to the assumption of fixed and known survey weights.

\section{Weighted Likelihood Adjustments}\label{sec: PL}
Another common approach to dealing with informatively sampled data in the survey realm is to incorporate the survey weights into the likelihood to get a ``pseudo-likelihood" of the form
\[\prod_{j\in\mathcal{S}, d\in\mathcal{D}} p(z_{jd}|\bm{\theta})^{w_{jd}},\]
where, as before, the weights $w_{jd}$ are inversely proportional to the probability of selection.
Inference is performed by solving the corresponding estimating equations
\begin{equation}\label{eq: estimating_eq} \sum_{j\in\mathcal{S}, d\in\mathcal{D}} w_{jd}\frac{\partial}{\partial\bm{\theta}}\log p(z_{jd}|\bm{\theta}) = 0
\end{equation}
and leads to design-consistent estimation of $\bm{\theta}$.

This strategy was introduced by \citet{bin83} and \citet{ski89}. As \citet{parker2019unit} detail, much recent work has been done in the survey literature to extend the pseudo-likelihood approach. These developments allow for the addition of random effects, hierarchical models, and the use of Bayesian inference.

In particular, \citet{sav16} show that, pseudo-likelihoods can reasonably be used in a Bayesian context. Given a prior distribution, $\pi(\bm{\theta})$, over $\bm{\theta}$, they prove that $L_1$ consistency is guaranteed for a pseudo-posterior of the form 
\[\hat{\pi}(\bm{\theta}|z_{jd}, w) \propto \left[\prod_{j\in\mathcal{S}, d\in\mathcal{D}} p(z_{jd}|\bm{\theta})^{\widetilde w_{jd}}\right]\pi(\bm{\theta})\]
for certain survey designs. In this case, the survey weights are normalized to sum to the sample size $\widetilde w_{jd} = \frac{w_{jd}}{\sum w_{jd}/n}$, so that the influence of each weight is on the order of the information in the sample.  They note that other formulations are possible, such as including a prior for the weights, or modeling them jointly with the response, but the ``plug-in" approach is the simplest and performs quite well. More recently, \citet{sav20} have extended this method to an even wider class of sampling designs. 

Pseudo-likelihoods have not been as widely adopted in the PS literature as the latent process approach of \citet{Diggle2010}, described in Section~\ref{sec: reg}. However, \citet{Zidek2014} implement a frequentist version of the pseudo-likelihood approach in the context of air quality monitoring. They analyze time series data for black smoke pollution in the UK, where the choice of sampling sites changed preferentially over several decades. Their approach to accounting for PS draws on the ``response-biased sampling" literature, such as \citet{scott2011}, as well as design-based survey inference. 

Given this official statistics perspective, they place greater emphasis on producing unbiased estimates compared to the geostatistical techniques outlined above, which focus more on prediction.  They further diverge from the geostatistical setup by assuming a finite population of possible sampling locations rather than a continuous domain. This formulation is more in line with the typical survey setting, which assumes a finite population.  They also modify the estimating equations in (\ref{eq: estimating_eq}) to allow for covariates.

As before, an obvious hurdle to translating any IS model to the PS setting is a lack of known weights in the latter case. To handle this, \citet{Zidek2014} use logistic regression to estimate the probability of selection for each site in the domain at each time point, $t$ based on the sample at time $t-1$, with some differences depending on whether the set of monitoring locations increases or decreases over time.

While the use of weighted likelihoods remains far less prevalent in the PS literature, the application in \citet{Zidek2014} suggests that many of the relevant developments in the survey literature could well be carried over to the PS problem. In fact, more recently, \cite{schliep2021correcting} have studied the use of weighted composite likelihoods in the context of spatial kriging under biased sampling schemes.

\section{Simulations}\label{sec: sim} 
To assess some of the approaches for handling PS data outlined so far, we carry out a set of simulation studies. We consider two ways of simulating PS data and compare the performance of each method in accounting for the preferential sample. 

\subsection{Scenario 1: Spatially Implicit Scenario}
In the first scenario, we begin by generating 1000 candidate points, $\mathbf{s}_i=(s_{1i}, s_{2i})',$ for ${i=1,\ldots,1000}$, uniformly over the unit square $\mathcal{D} = [0,1]\times[0,1]$. To take a preferential sample, we thin these candidate points with probability proportional to the response at each point. We keep each point with probability
\begin{equation}
p(\mathbf s_i) = (1-(s_{1i}-0.5)^2 - (s_{2i}-0.5)^2)^8.\label{pdef}
\end{equation}
This selects points closer to the center of the domain with higher probability than those near the edges of the domain.
 We then generate values of a response, $z_i$, from the model 
\[z(\mathbf s_i) = (5,2)'\mathbf s_{i} + 2\widetilde{p}(\mathbf s_i) + \varepsilon_{i},\]
where $\varepsilon_{i} \sim N(0,.5)$ and $\widetilde{p}(\mathbf s_i)$ is probability of selection as defined in (\ref{pdef}), now centered and scaled. Including the probability of selection in the response introduces a dependency between the sampling scheme and the response.

\begin{enumerate}
    \item As a baseline, we fit a Bayesian linear regression model that does not attempt to correct for the PS scheme
     \begin{align*}
      \mathbf Z | \boldsymbol\beta,\sigma_z &\propto \prod_{i\in\mathcal{S}} N(z_i|\boldsymbol s_i'\boldsymbol \beta,\sigma_z)\\
      \boldsymbol \beta &\sim N_2(0,\sigma^2_\beta \mathbf I_2)\\
      \sigma_z &\sim \text{Cauchy}^+(0,10),
     \end{align*} 
     where $\mathbf{s}_i$ is the two-dimensional coordinate vector for the $i$-th sample point and $\boldsymbol {\beta}$ is a vector of  their associated regression coefficients. For our simulations, we set $\sigma^2_\beta =\sqrt{10}$. This can be seen as a pseudo-likelihood model with unit weights, hence we refer to it as ``unweighted" (UW).
    \item We then fit a set of weighted pseudo-likelihood models similar to those outlined in Section~\ref{sec: PL}. These are specified identically to the UW model, but with the addition of scaled weights, $\widetilde{w}_i=\widetilde w(\mathbf{s}_i)$, that correct for the underlying selection probability, so that
    \begin{align*}
      \mathbf Z | \boldsymbol\beta,\sigma_z &\propto \prod_{i\in\mathcal{S}} [N(z_i|\boldsymbol s_i'\boldsymbol \beta,\sigma_z)]^{\widetilde{w}_i}\\
      \boldsymbol \beta &\sim N_2(0,\sigma^2_\beta \mathbf I_2)\\
      \sigma_z &\sim \text{Cauchy}^+(0,10).
    \end{align*}
        As before, we set $\sigma_\beta^2=\sqrt{10}$ in our simulations.
        
        We compare the performance of two different schemes for defining the weights. First, as a baseline for this scenario, we take the true, known probability of selection, $p(\mathbf s_i)$ at each sample point $\mathbf s_i$ and set $w_i=w(\mathbf s_i)=1/p(\mathbf s_i)$, then rescale so that these weights sum to the sample size. We refer to this model as the pseudo-likelihood known weights (PKW) model.
    
        In practice, the true weights are not typically known, so for a comparison that might actually be used in practice,  we obtain a kernel density estimate of the probability of selection using a Gaussian kernel via the \texttt{MASS} package with default bandwidth \citep{MASS}. As before, we rescale these weights so that they sum to the sample size. We refer to this model as the pseudo-likelihood, estimated weights (PEW) model. 
        
    \item Lastly, we fit a hierarchical model as defined in \cite{Pati2011} and described in Section~\ref{sec: reg}, where a latent GP is shared between the response and the point process. We refer to this model as the PRD model. Fitting this model requires discretizing the domain, and we follow the authors' recommendations for an equally spaced grid of 225 knots on $[-0.2, 1.2]^2$ over a square grid of $41\times 41$ points, which ensures the grid spacings are chosen to be no larger than the standard deviation of the kernel in the convolution representation. 
\end{enumerate}

The UW, PKW, and PEW models are fit using Hamiltonian Monte Carlo via Stan. Each of these models is run for 5500 iterations with the first 1000 burn-in steps discarded. The PRD model is estimated using 60,000 iterations with 10,000 burn-in steps discarded. 
 We repeat each of the simulations 100 times with an initial candidate set size 
of 1,000 locations each time. Visual inspection of the trace plots as well as effective sample size of the sample chains indicated no lack of convergence for any of the models.

\begin{figure}
    \begin{center}
        \includegraphics[width=5in]{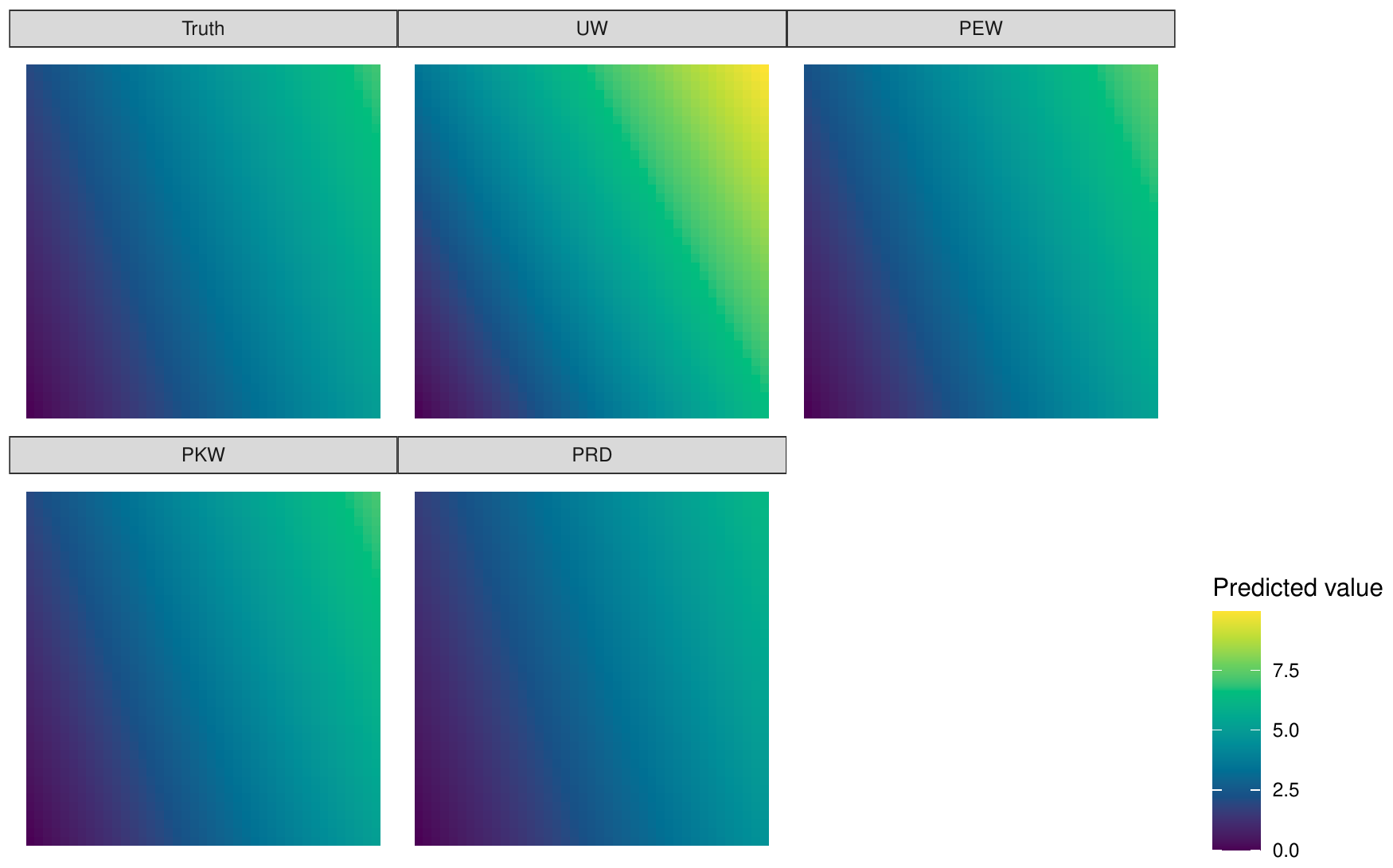}
    \end{center}               
         \caption{Plots of the predicted surfaces for each of the four models in the spatially implicit simulation scenario along with the true surface for comparison.\label{fig:sim1_surfaces}}
\end{figure}

The mean estimate, credible interval (CI) coverage rate, and mean CI width for each parameter are reported in Table~\ref{tab:sim1_params}. Overall MSE and bias for the predicted surfaces are reported in Table~\ref{tab:sim1_results} and posterior mean surface plots are provided in Figure~\ref{fig:sim1_surfaces}. In terms of parameter estimates, all three PS models are able to greatly reduce the bias compared to the UW model, with the PL approaches performing best in this case. In terms of accuracy of uncertainty estimates, as assessed by the CI coverage rate, again the PS models are able to greatly improve upon the UW model. However, for $\beta_0$ specifically, the PRD model has coverage much closer to the nominal level than the PL models. Looking at predictive ability (Table~\ref{tab:sim1_results}), the unweighted model grossly underperforms each of the corrective models and shows substantial bias. The corrective models all show greatly reduced bias, where the best results in terms of MSE are achieved with the PKW model. However, since full knowledge of the underlying sample location intensity surface is impractical in reality, the PEW model would be the most reasonable in practice. 

\begin{table}[ht]
    \centering
     \begin{tabular}{||c|ccc|ccc||}\hline
       && $\hat\beta_1$  & & & $\hat\beta_2$ &  \\\hline \hline
Model& Mean  &  CI coverage & Mean CI width       & Mean    &  CI coverage   & Mean CI width\\\hline 
UW  & 6.408 &   2\%        &   1.311             & 3.520   &  0\%          &   1.314       \\
PEW & 5.278 &   73\%       &   1.032             & 2.258   &  83\%         &   1.028        \\
PKW & 5.101 &   61\%       &   0.983             & 1.989   &  86\%         &   0.976        \\
PRD & 4.604 &   89\%       &   2.174             & 1.676   &  98\%         &   2.167\\ \hline        
    \end{tabular}
    \caption{Average of parameter estimates, as well as 90\% CI coverage probabilities and average CI widths, for each of the four models in the spatially implicit scenario simulation. The true parameter values are $\beta_1=5$ and $\beta_2=2.$}
    \label{tab:sim1_params}
\end{table}

\begin{table}[ht]
    \centering
     \begin{tabular}{||c|c|c||}\hline
     Model       & MSE & Mean Abs. Bias\\ \hline\hline 
     \hline 
     UW  & 2.551 & 1.464\\
     PEW & 0.121 & 0.268        \\
     PKW & 0.081 & 0.046        \\
     PRD        & 0.380 & 0.360 \\\hline
    \end{tabular}
    \caption{MSE and mean absolute bias for the posterior mean predictive surface produced by each of the four models in the spatially implicit scenario simulation.}
    \label{tab:sim1_results}
\end{table}

The runtime of unweighted and pseudo-likelihood models are quite similar. Taking the unweighted as a baseline, the relative runtime for PEW to fit is only 1.019 times longer, while for PKW, it is 0.957 longer. The relative runtime for PRD is several orders of magnitude longer, but a more optimized implementation would likely narrow this discrepancy.

\subsection{Simulation 2: Spatially Explicit Scenario}
In the second scenario, we begin by simulating an intensity surface, $p(s)$, from a Gaussian process with a squared exponential kernel using the covariance function
\[c(s_i, s_j) = \exp\left(-2\sum_{d=1}^D(s_{di}-s_{dj})^2\right).\]
Specifically, we let $\alpha=1$, and $\rho=0.5$.

\begin{figure}
    \begin{center}
        \includegraphics[width=5in]{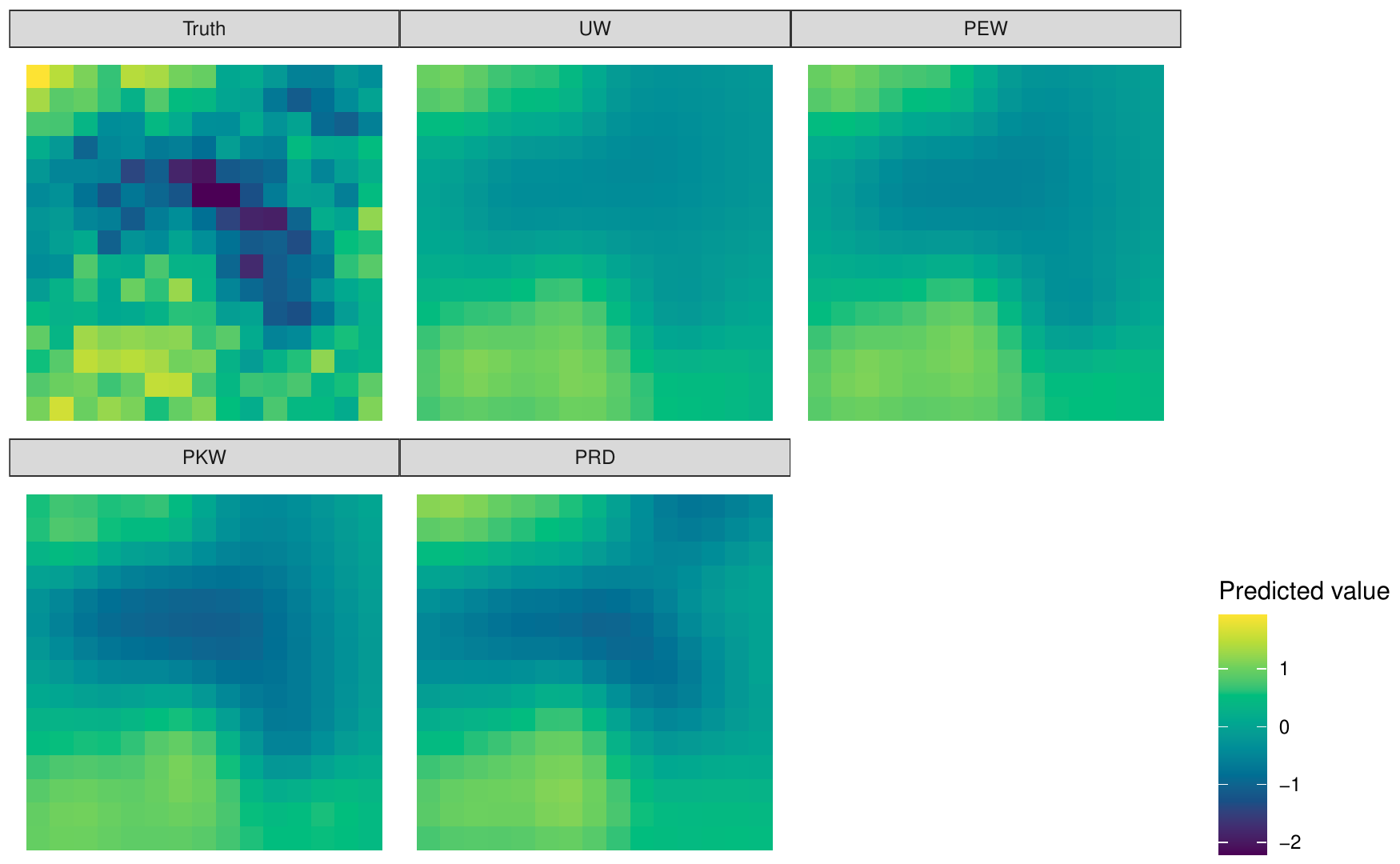}
    \end{center}               
         \caption{\baselineskip=10pt Plots of the posterior mean predicted surfaces for each of the four models in the spatially explicit simulation scenario along with the true surface for comparison.\label{fig:sim2_surfaces}}
\end{figure}

We then implement a PS scheme by selecting sample locations from an inhomogenous Poisson point process using the generated intensity function. Finally, observed values, $Z(s)$, are generated by adding iid Gaussian noise to the intensity evaluated at the observed locations. In other words, $Z(s) = p(s) + \epsilon(s). $ Thus, this approach is in essence generating data from the model specified by \cite{Diggle2010}. The resulting true surface is reproduced in Figure~\ref{fig:sim2_surfaces}.

We fit the same three models as in Simulation 1, but now with a spatial process for the mean rather than a linear combination of spatial covariates. We choose to use a basis expansion representation of the spatial process. Thus, for this scenario, the UW model is given as
    \begin{align*}
        Z(s) | \bm{\eta}, \sigma^2_z & \sim \hbox{N}(p(s), \sigma^2_z) \\
        p(s) &= \sum_{k=1}^K \phi_k(s)\eta_k \\
        \eta_k & \sim \hbox{N}(0, \lambda_k \tau), \; k=1,\ldots,K \\
        \lambda_k &\sim \hbox{Cauchy}^+(0,1), \; k=1,\ldots,K \\
        \tau & \sim \hbox{Cauchy}^+(0,1),
    \end{align*}
where $\phi_k(s)$ is the value of the $k$th basis function evaluated at location $s.$ An initially large number of basis functions is selected automatically at two resolutions using the \texttt{FRK} package \citep{zammit2021frk}. A horseshoe prior \citep{carvalho2010horseshoe} is then placed on $\eta_k$ in order to provide shrinkage for the initially large number of basis functions.

Similar to the first simulation scenario, we also compare to weighted pseudo-likelihood versions of this model. As before, we fit a pseudo-likelihood model with known weights (PKW) and a pseudo-likelihood model with weights constructed using a kernel density estimator (PEW) with the same settings as in the first simulation scenario. In addition, we compare to the method used by \cite{Pati2011} (PRD). 

The results of this simulation are summarized in Table~\ref{tab:sim2_results}. In this scenario, the PRD model outperforms the pseudo-likelihood models both in terms of MSE and mean absolute bias. This is to be expected here, as the data generating model is a case of the PRD model. However, each of the corrective models is able to reduce the MSE and bias relative to the UW model. Finally, the difference in MSE and mean absolute bias between the PEW and PKW models is less pronounced than in Simulation 1. 

Taking UW as a baseline, the relative runtime for PEW is 0.913. For PKW it is 0.901, while for PRD it is 49.340.
\begin{table}[h]
    \centering 
     \begin{tabular}{||c|c|c|c||}\hline
     Model         & MSE & Mean Abs. Bias \\    
     \hline\hline
     UW           & 0.382 & 0.418      \\      
     PEW          & 0.332 & 0.376      \\     
     PKW          & 0.304 & 0.341      \\    
     PRD          & 0.245 & 0.312      \\\hline
    \end{tabular}
    \caption{MSE and mean absolute bias for the posterior mean predictive surface produced by each of the four models in the spatially explicit scenario simulation.}
    \label{tab:sim2_results}
\end{table}

\section{Application to  Heavy Metal Biomonitoring Data}\label{sec: examp}

\begin{figure}
\begin{center}
  \includegraphics[width=3in]{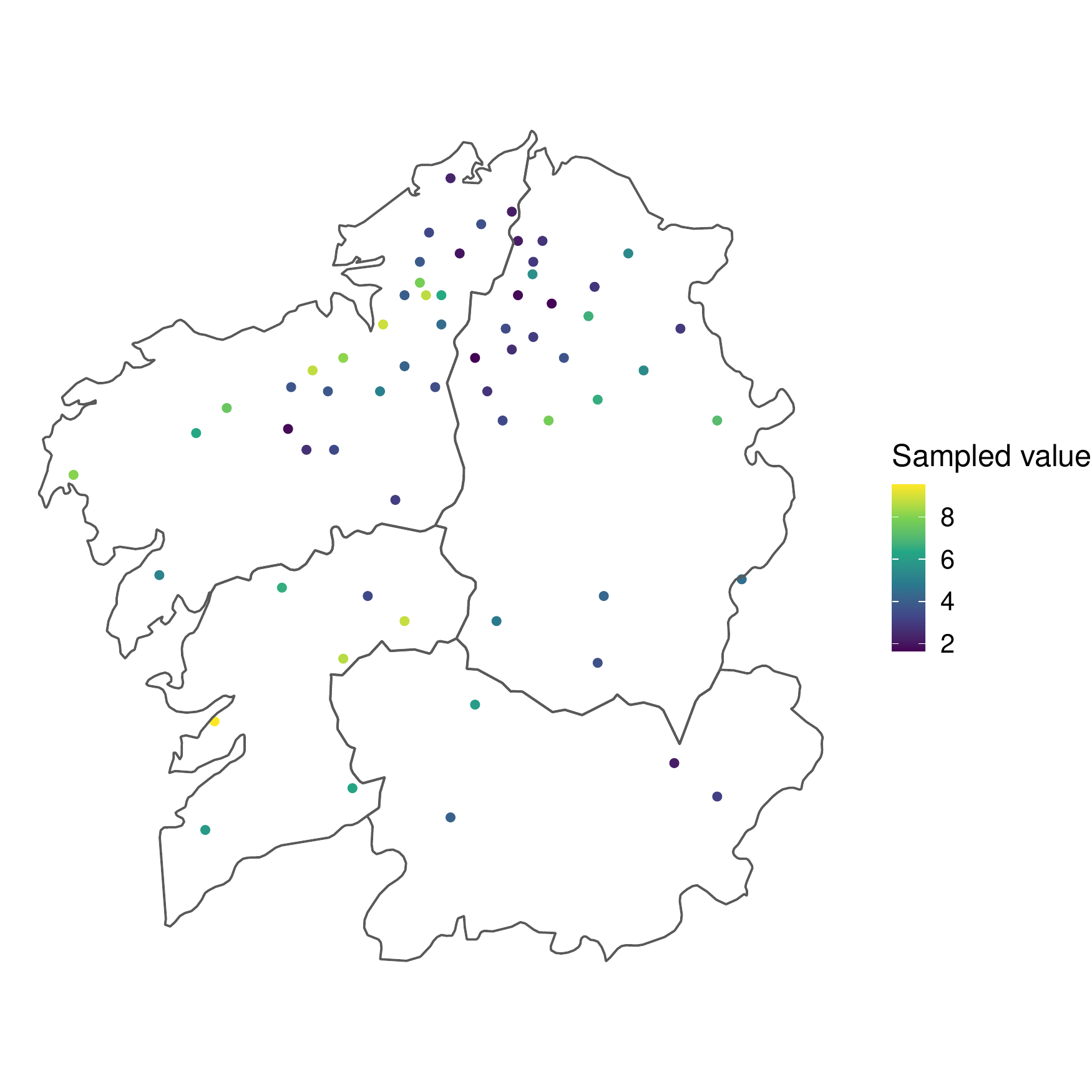}
\end{center}
  \caption{Plot of the sampling locations for the heavy metal bio-monitoring data in Galicia under a preferential scheme. \label{fig:galicia_pref_sample}}
\end{figure}

\begin{figure}
\begin{center}
  \includegraphics[width=3in]{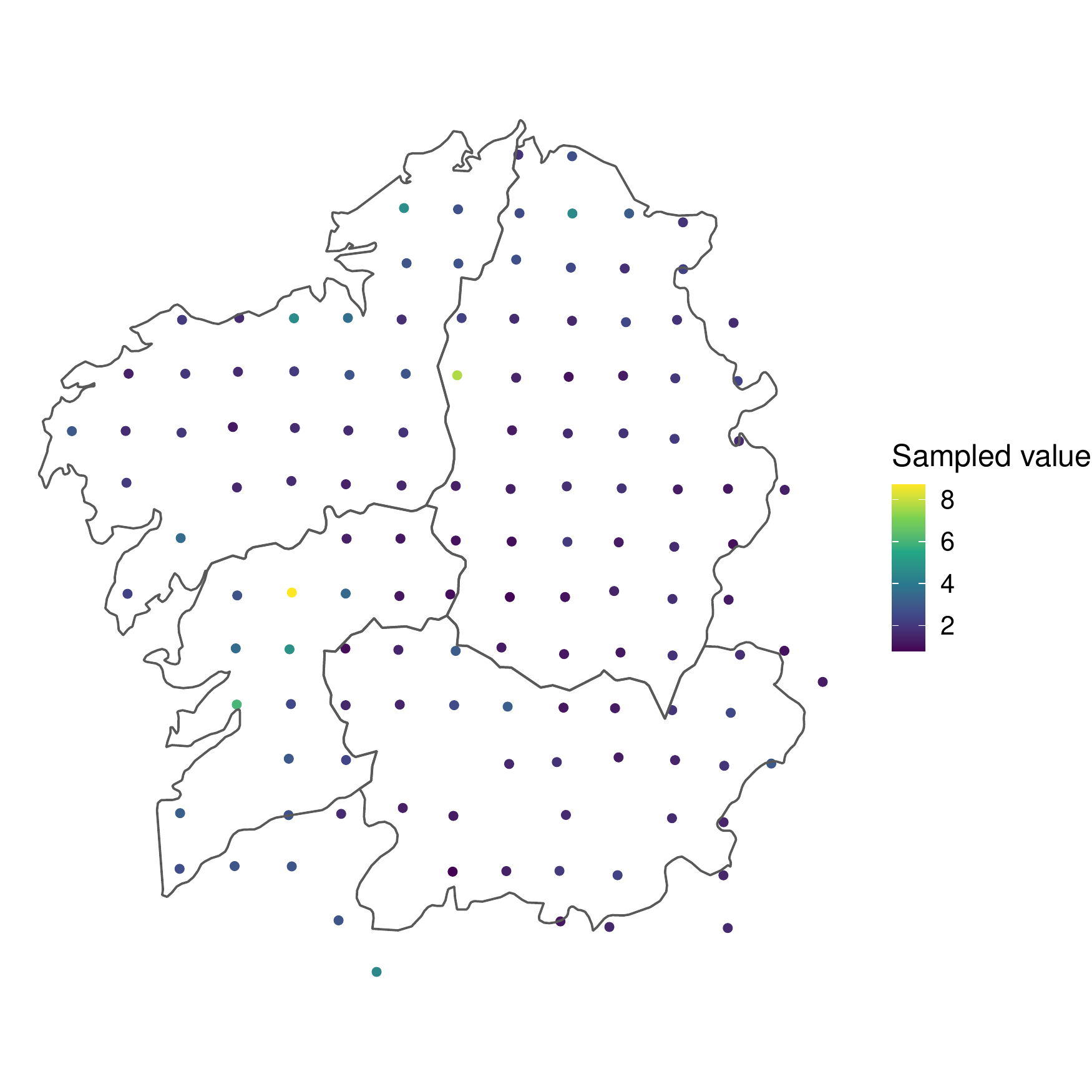}
\end{center}
  \caption{Plots of the sampling locations for the heavy metal bio-monitoring data in Galicia under a non-preferential, regular lattice sample. The boundary is only approximate, so some points appear to fall outside of it, but these points do not affect the analysis.  \label{fig:galicia_nonpref_sample}}
\end{figure}

To illustrate the different approaches in practice, we fit three of the models from the second simulation scenario to the heavy metal biomonitoring data from  Galicia, Spain analyzed in \cite{Diggle2010}. The response is lead concentrations in micorgrams per gram dry weight of moss. The data come from two surveys only a few years apart. The 1997 data (63 sample points) uses a PS scheme, where sites with ``large gradients" were more likely to be sampled. Figure~\ref{fig:galicia_pref_sample} shows the sampled locations, with a clear concentration in the northern part of the region. The 2000 data (132 sample points) uses a regular lattice sample (i.e. not preferential) over the same domain (Figure~\ref{fig:galicia_nonpref_sample}). One question of interest could be whether a difference in mean response across the two samples is attributable to a true change over time, or only to the difference in sampling scheme. 


\begin{figure}
    \begin{center}
        \includegraphics[width=5in]{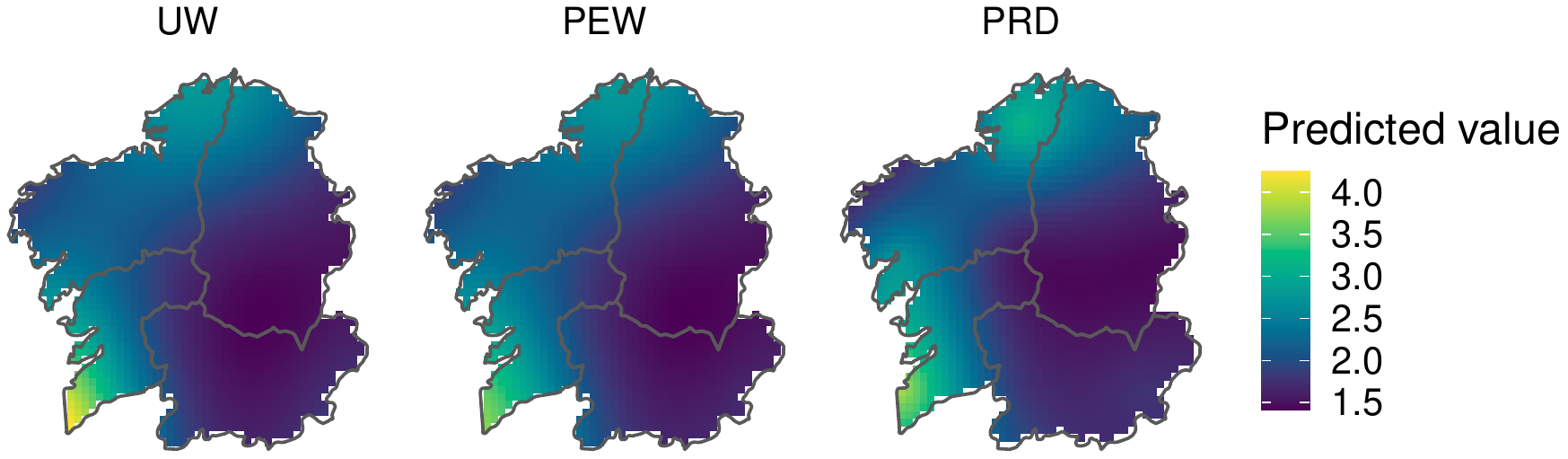}
    \end{center}               
         \caption{\baselineskip=10pt Predicted surfaces over Galicia for each model fit to the heavy metal bio-monitoring data under the non-preferential, regular lattice, sampling scheme.\label{fig:real_nonpref_surfaces}}
\end{figure}

\begin{figure}
    \begin{center}
        \includegraphics[width=5in]{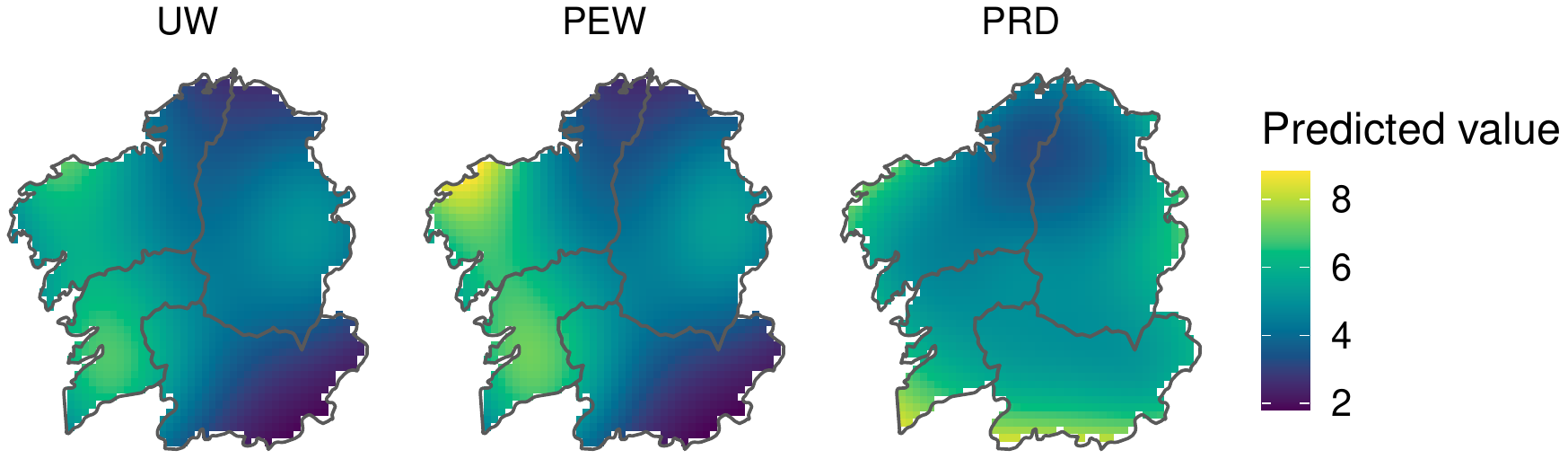}
    \end{center}               
         \caption{\baselineskip=10pt Predicted surfaces over Galicia for each model fit to the heavy metal bio-monitoring data under the preferential sampling scheme.
         \label{fig:real_pref_surfaces}}
\end{figure}

With real data, we have no way of knowing the true surface, and this rules out fitting the PKW model of Section~\ref{sec: sim}. However, by fitting models that do and do not assume the presence of PS to both datasets, we can evaluate the degree to which the model estimates differ relative to each other. The fit for each of the models to the non-preferential data is shown in Figure~\ref{fig:real_nonpref_surfaces}, while the fit for the preferential data is shown in Figure~\ref{fig:real_pref_surfaces}. As expected, all three models yield similar estimates when the data is sampled in a non-preferential manner. For the preferentially sampled data, there is much more variation in the estimates across models. The PEW model seems to give larger predictions than the UW model along the western edge, where predictions are generally higher than average. Interestingly, the PRD model results in high predicted heavy metals along the southeast edge, contradictory to both other models, as well as the non-preferential sample. In this case, there is no true baseline to compare to, however, we might expect the true surface to be somewhat similar in the non-preferential and the preferential sample, since they both sample the same geographic domain, although at different times.

\section{Discussion}\label{sec: disc}
The problems of PS and IS have recently undergone extensive research, with many proposed methods within each domain. Although the overlap between the two problems is significant, in large part, the research has evolved down two independent tracks. In other words, little attention has been devoted to the explicit connections between these two related problems.

In this paper we review many of the existing solutions to each of these problems and note the important differences in the modeling techniques. In addition, we provide a comprehensive review and comparison of the methods for addressing PS and note that a more extensive review of methods for addressing IS can be found in \citet{parker2019unit}.

Broadly speaking, there are several approaches for handling the problems of PS and IS, including use of informative covariates in the model, regressing on the selection process, and weighted likelihood adjustments. Even though we demonstrate that there is significant overlap between the approaches used to handle each problem, one important distinction is that in the context of IS the weights are typically known from the sampling design, which is not the case for PS. In contrast, in the context of PS, the weights need to be estimated if they are to be used in a model.

To demonstrate, compare, and contrast the different methods in the context of PS, we provide a comprehensive simulation study. Notably, as expected, we find that the methods accounting for PS outperform the unweighted approach. This corroborates the findings in the IS context (e.g., see \citet{parker2019unit}). Finally, through an application to heavy metal monitoring data, we illustrate the difference in analyses that account for PS relative to analyses that ignore PS.

There are several opportunities for future research. For example, methods in PS can be adapted to IS and vice versa. To this end, this work is meant to bridge the gap between the IS and PS communities and promote future research in both areas.

\section*{Acknowledgements}

This research was partially supported by the U.S.~National Science Foundation (NSF) under NSF grant SES-1853096. 
This article is released to inform interested parties of ongoing research and to encourage discussion. The views expressed on statistical issues are those of the authors and not those of the NSF or U.S. Census Bureau.

\clearpage

\bibliography{myrefs}

\begin{thebibliography}{33}
\newcommand{\enquote}[1]{``#1''}
\expandafter\ifx\csname natexlab\endcsname\relax\def\natexlab#1{#1}\fi

\bibitem[\protect\citename{Binder, }1983]{bin83}
Binder, D.~A. (1983).
\newblock \enquote{On the variances of asymptotically normal estimators from
  complex surveys.}
\newblock {\em International Statistical Review\/}, 51, 3, 279--292.

\bibitem[\protect\citename{Carvalho et~al., }2010]{carvalho2010horseshoe}
Carvalho, C.~M., Polson, N.~G., and Scott, J.~G. (2010).
\newblock \enquote{The horseshoe estimator for sparse signals.}
\newblock {\em Biometrika\/}, 97, 2, 465--480.

\bibitem[\protect\citename{Conn et~al., }2017]{conn17}
Conn, P.~B., Thorson, J.~T., and Johnson, D.~S. (2017).
\newblock \enquote{Confronting preferential sampling when analysing population
  distributions: diagnosis and model-based triage.}
\newblock {\em Methods in Ecology and Evolution\/}, 8, 11, 1535--1546.

\bibitem[\protect\citename{da~Silva~Ferreira and Gamerman,
  }2015]{da2015optimal}
da~Silva~Ferreira, G. and Gamerman, D. (2015).
\newblock \enquote{Optimal design in geostatistics under preferential
  sampling.}
\newblock {\em Bayesian Analysis\/}, 10, 3, 711--735.

\bibitem[\protect\citename{Diggle et~al., }2010]{Diggle2010}
Diggle, P.~J., Menezes, R., and li~Su, T. (2010).
\newblock \enquote{Geostatistical inference under preferential sampling.}
\newblock {\em Journal of the Royal Statistical Society: Series C (Applied
  Statistics)\/}, 59, 2, 191--232.

\bibitem[\protect\citename{Dinsdale and Salibian-Barrera,
  }2019]{dinsdale2019methods}
Dinsdale, D. and Salibian-Barrera, M. (2019).
\newblock \enquote{Methods for preferential sampling in geostatistics.}
\newblock {\em Journal of the Royal Statistical Society: Series C (Applied
  Statistics)\/}, 68, 1, 181--198.

\bibitem[\protect\citename{Firth and Bennett, }1998]{firth98}
Firth, D. and Bennett, K. (1998).
\newblock \enquote{Robust models in probability sampling.}
\newblock {\em Journal of the Royal Statistical Society: Series B (Statistical
  Methodology)\/}, 60, 1, 3--21.

\bibitem[\protect\citename{Gelfand et~al., }2012]{gelfand12}
Gelfand, A.~E., Sahu, S.~K., and Holland, D.~M. (2012).
\newblock \enquote{On the effect of preferential sampling in spatial
  prediction.}
\newblock {\em Environmetrics\/}, 23, 7, 565--578.

\bibitem[\protect\citename{Gelfand and Shirota, }2019]{gelfand19}
Gelfand, A.~E. and Shirota, S. (2019).
\newblock \enquote{Preferential sampling for presence/absence data and for
  fusion of presence/absence data with presence-only data.}
\newblock {\em Ecological Monographs\/}, 89, 3, e01372.

\bibitem[\protect\citename{Gelman and Little, }1997]{gelman_little1997}
Gelman, A. and Little, T.~C. (1997).
\newblock \enquote{Poststratification into many categories using hierarchical
  logistic regression.}

\bibitem[\protect\citename{Grantham et~al., }2018]{grantham2018spatial}
Grantham, N.~S., Reich, B.~J., Liu, Y., and Chang, H.~H. (2018).
\newblock \enquote{Spatial regression with an informatively missing covariate:
  Application to mapping fine particulate matter.}
\newblock {\em Environmetrics\/}, 29, 4, e2499.

\bibitem[\protect\citename{Irvine et~al., }2018]{irvine2018occupancy}
Irvine, K.~M., Rodhouse, T.~J., Wright, W.~J., and Olsen, A.~R. (2018).
\newblock \enquote{Occupancy modeling species--environment relationships with
  non-ignorable survey designs.}
\newblock {\em Ecological Applications\/}, 28, 6, 1616--1625.

\bibitem[\protect\citename{Jiao et~al., }2019]{jiao2019bayesian}
Jiao, J., Hu, G., and Yan, J. (2019).
\newblock \enquote{A Bayesian Joint Model for Spatial Point Processes with
  Application to Basketball Shot Chart.}
\newblock {\em arXiv preprint arXiv:1908.05745\/}.

\bibitem[\protect\citename{Little, }2012]{little2012}
Little, R.~J. (2012).
\newblock \enquote{Calibrated Bayes, an alternative inferential paradigm for
  official statistics.}
\newblock {\em Journal of Official Statistics\/}, 28, 3, 309.

\bibitem[\protect\citename{Little, }1993]{little1993}
Little, R. J.~A. (1993).
\newblock \enquote{Post-Stratification: A Modeler's Perspective.}
\newblock {\em Journal of the American Statistical Association\/}, 88, 423,
  1001--1012.

\bibitem[\protect\citename{M{\o}ller et~al., }1998]{moller1998log}
M{\o}ller, J., Syversveen, A.~R., and Waagepetersen, R.~P. (1998).
\newblock \enquote{Log Gaussian Cox processes.}
\newblock {\em Scandinavian Journal of Statistics\/}, 25, 3, 451--482.

\bibitem[\protect\citename{Park et~al., }2006]{park06}
Park, D.~K., Gelman, A., and Bafumi, J. (2006).
\newblock {\em 11. State-Level Opinions from National Surveys:
  Poststratification Using Multilevel Logistic Regression\/},  209--228.
\newblock Stanford University Press.

\bibitem[\protect\citename{Parker et~al., }2019]{parker2019unit}
Parker, P.~A., Janicki, R., and Holan, S.~H. (2019).
\newblock \enquote{Unit level modeling of survey data for small area estimation
  under informative sampling: A comprehensive overview with extensions.}
\newblock {\em arXiv preprint arXiv:1908.10488\/}.

\bibitem[\protect\citename{Pati et~al., }2011]{Pati2011}
Pati, D., Reich, B.~J., and Dunson, D.~B. (2011).
\newblock \enquote{Bayesian geostatistical modelling with informative sampling
  locations.}
\newblock {\em Biometrika\/}, 98, 1, 35--48.

\bibitem[\protect\citename{Pennino et~al., }2019]{pennino2019accounting}
Pennino, M.~G., Paradinas, I., Illian, J.~B., Mu{\~n}oz, F., Bellido, J.~M.,
  L{\'o}pez-Qu{\'\i}lez, A., and Conesa, D. (2019).
\newblock \enquote{Accounting for preferential sampling in species distribution
  models.}
\newblock {\em Ecology and Evolution\/}, 9, 1, 653--663.

\bibitem[\protect\citename{Pfeffermann and Sverchkov, }2007]{pfe07}
Pfeffermann, D. and Sverchkov, M. (2007).
\newblock \enquote{Small-area estimation under informative probability sampling
  of areas and within the selected areas.}
\newblock {\em Journal of the American Statistical Association\/}, 102, 480,
  1427--1439.

\bibitem[\protect\citename{Savitsky and Toth, }2016]{sav16}
Savitsky, T.~D. and Toth, D. (2016).
\newblock \enquote{{Bayesian estimation under informative sampling}.}
\newblock {\em Electronic Journal of Statistics\/}, 10, 1, 1677 -- 1708.

\bibitem[\protect\citename{Schliep et~al., }2021]{schliep2021correcting}
Schliep, E.~M., Wikle, C.~K., and Daw, R. (2021).
\newblock \enquote{Correcting spatial Gaussian process parameter and prediction
  variance estimation under informative sampling.}
\newblock {\em arXiv preprint arXiv:2108.12354\/}.

\bibitem[\protect\citename{Scott and Wild, }2011]{scott2011}
Scott, A.~J. and Wild, C.~J. (2011).
\newblock \enquote{Fitting regression models with response-biased samples.}
\newblock {\em The Canadian Journal of Statistics / La Revue Canadienne de
  Statistique\/}, 39, 3, 519--536.

\bibitem[\protect\citename{Si et~al., }2015]{si15}
Si, Y., Pillai, N.~S., Gelman, A., et~al. (2015).
\newblock \enquote{{B}ayesian nonparametric weighted sampling inference.}
\newblock {\em Bayesian Analysis\/}, 10, 3, 605--625.

\bibitem[\protect\citename{Skinner, }1989]{ski89}
Skinner, C.~J. (1989).
\newblock \enquote{Domain means, regression and multivariate analysis.}
\newblock In {\em Analysis of Complex Surveys\/}, eds. C.~J. Skinner, D.~Holt,
  and T.~M.~F. Smith, chap.~2,  80 -- 84. Chichester: Wiley.

\bibitem[\protect\citename{Vandendijck et~al., }2016]{van16}
Vandendijck, Y., Faes, C., Kirby, R.~S., Lawson, A., and Hens, N. (2016).
\newblock \enquote{Model-based inference for small area estimation with
  sampling weights.}
\newblock {\em Spatial Statistics\/}, 18, 455--473.

\bibitem[\protect\citename{Venables and Ripley, }2002]{MASS}
Venables, W.~N. and Ripley, B.~D. (2002).
\newblock {\em Modern Applied Statistics with S\/}.
\newblock 4th ed. New York: Springer.
\newblock ISBN 0-387-95457-0.

\bibitem[\protect\citename{Watson, }2021]{watson2021perceptron}
Watson, J. (2021).
\newblock \enquote{A perceptron for detecting the preferential sampling of
  locations and times chosen to monitor a spatio-temporal process.}
\newblock {\em Spatial Statistics\/}, 43, 100500.

\bibitem[\protect\citename{Williams and Savitsky, }2020]{sav20}
Williams, M.~R. and Savitsky, T.~D. (2020).
\newblock \enquote{{Bayesian Estimation Under Informative Sampling with
  Unattenuated Dependence}.}
\newblock {\em Bayesian Analysis\/}, 15, 1, 57 -- 77.

\bibitem[\protect\citename{Zammit-Mangion and Cressie, }2021]{zammit2021frk}
Zammit-Mangion, A. and Cressie, N. (2021).
\newblock \enquote{FRK: An R package for spatial and spatio-temporal prediction
  with large datasets.}
\newblock {\em Journal of Statistical Software\/}, 98, 1--48.

\bibitem[\protect\citename{Zheng and Little, }2003]{zhe03}
Zheng, H. and Little, R.~J. (2003).
\newblock \enquote{Penalized spline model-based estimation of the finite
  populations total from probability-proportional-to-size samples.}
\newblock {\em Journal of Official Statistics\/}, 19, 2, 99.

\bibitem[\protect\citename{Zidek et~al., }2014]{Zidek2014}
Zidek, J.~V., Shaddick, G., and Taylor, C.~G. (2014).
\newblock \enquote{Reducing estimation bias in adaptively changing monitoring
  networks with preferential site selection.}
\newblock {\em The Annals of Applied Statistics\/}, 8, 3, 1640--1670.

\end{thebibliography}
\bibliographystyle{jasa}

\end{document}